\documentclass[iop,aps]{revtex4} 
\usepackage{graphicx}

\begin{document}

\title{Ab initio simulations of $H_2$ in Li-doped carbon nanotube systems}
\author{Abdenour Sabir, Wenchang Lu, Christopher Roland, Jerzy Bernholc}

\affiliation{Department of Physics, North Carolina State
University, Raleigh, North Carolina 27695}

\begin{abstract}

Because of their unique structure, it has been proposed that carbon nanotube ropes may
well provide an ideal container for the storage of molecular hydrogen. Indeed, there
has been some experimental evidence of enhanced hydrogen uptake in doped Li and
other alkali metal systems [Chen {\it et al}, Science {\bf 285}, 91 (1999)]. We have therefore 
addressed this issue of hydrogen storage in Li-doped graphite and carbon nanotube systems
theoretically with {\it ab initio} simulations. Our results find no evidence for
such enhanced storage, based on the induced structural changes. In addition, we have
also investigated the diffusion barriers for hydrogen to enter into nanotube interiors,
both in the presence and absence of topological defects. Even if nanotube
interiors are made accessible, the hydrogen uptake remains modest, {\it i.e.}, less
than 3.5 wt$\%$. Mechanically or chemically processing nanotubes is therefore not
likely to lead to greatly increased hydrogen storage.
\pacs{}
\end{abstract}

\maketitle

\smallskip
\centerline
{\bf I. Introduction}

\bigskip

It has long been recognized that molecular hydrogen is an almost ideal, environmentally
friendly fuel \cite{Schlapbach}. However, to be truly useful, the technological ability to store hydrogen
molecules must be considerably improved \cite {Dress}. For these purposes, metal hydridization 
remains a distinct possibility, but is not a preferred solution because of the associated expense
and weight. Carbon nanotubes ropes and other carbon-based systems have been
proposed as alternative hydrogen storage materials. The key feature that makes nanotubes so 
potentially useful as a storage material is their structure. Carbon nanotubes \cite {Ijima} 
are formed when a graphene sheet
is curled into a cylinder and the carbon atoms joined seamlessly to each other. During
growth, individual nanotubes typically aggregate into bundles, termed {\it ropes}, which
consist of potentially hundreds of nanotubes held together by van der Waals forces. These
ropes are expected to display superior hydrogen uptake, provided that the hydrogen molecules
can intercalate both into the channels between the nanotubes, and into the interior of the
nanotube themselves. Based on these expectations, initial experiments of hydrogen storage
in nanotubes were very encouraging: for instance, work by Dillon {\it et
al} \cite {Dillon} reported an uptake of 7 weight (wt) percent of $H_2$ at 133 K and 0.04 MPa; 
Ye {\it and coworkers} \cite {Ye and al} used high purity nanotubes and obtained 8.25 wt $\%$ at 80K 
and 7 Mpa. Unfortunately, subsequent experiments \cite {Strobel} failed to reproduce these 
results for a variety of reasons, so that the resulting controversy has not yet been
fully resolved. More recently, another interesting alternative for enhanced hydrogen storage was given 
by Chen and coworkers \cite {Chen}, whose experiments claimed to achieve up to 20 wt $\%$ 
uptake in Li-doped carbon nanotube systems, and up to 14 wt $\%$ for K-doped nanotube systems. 
These numbers are impressive and compare favorably with those of gasoline or diesel, which contain 
the equivalent of 17 wt $\%$. This new approach has the additional advantage of operating at ambient pressures and modest temperature.

Motivated by these experiments, we have re-examined the question of hydrogen storage
in carbon-based systems with {\it ab initio} simulations. 
In particular, we focus on the issue of enhanced hydrogen adsorption in Li-intercalated 
graphite and carbon nanotube systems. Unfortunately, in agreement with the most recent 
experimental results \cite {Ritschel} and other simulations of hydrogen/nanotube systems, 
we find no evidence or mechanisms for enhanced hydrogen storage. In addition, since the key 
to enhanced hydrogen uptake in nanotube ropes is predicated on the ability of molecules
to move in and out of the nanotube interiors, we have studied the diffusion barriers for 
hydrogen molecules to enter nanotube interiors. Both {\it pristine} and {\it defective} 
nanotubes have been considered. The latter is particular important, because defects may readily be
induced in nanotube side-walls via mechanical \cite {Liu} and chemical means
\cite {Pajayan} . Indeed, both experiments\cite{Gao}and theory \cite{Chris-Vincent} show that 
Li-uptake in nanotube systems, for battery applications, is increased substantially 
when the nanotubes are first subjected to an intense ball-milling treatment. Our calculations 
show, that even when very large defects are present, there are substantial diffusion barriers
present so that only moderate increases in the hydrogen uptake may be expected.

\bigskip

\centerline
{\bf II. Results}

\smallskip
\noindent 
{\bf (a) Methodology} 

To investigate the issue of hydrogen storage in Li-doped carbon-based systems, we
performed {\it ab initio} density functional theory (DFT) based calculations using a
real-space code with multigrid acceleration\cite{Briggs}. 
Nonlocal and norm-conserving pseudopotentials of the Troullier-Martins form\cite{Troullier} 
were employed using the standard Kleinman-Bylander form\cite{Kleinman}. Calculations were 
carried out with a grid spacing corresponding to an effective cutoff of 63 Ry with 12 k-points. To calculate the diffusion barriers, we mapped out the point-by-point diffusion pathways, while keeping the orientation of the hydrogen molecules fixed. Calculations were carried out both with the
local density approximation (LDA) and with generalized gradient corrections
(GGA) of Perdew-Burke-Ernzerhof (PBE)\cite{Ernzerhof}. For this
issues related to hydrogen storage, both methods give very similar results, and hence we shall
mostly quote only the LDA results, noting differences as they arise.
We now turn to the results of our calculations. Since carbon nanotubes are structurally intimately
related to graphene sheets, we first consider hydrogen molecules in graphite.

\smallskip
\noindent  
{\bf (b) $H_2$ in doped graphite}

\begin{figure}[htbc]
\begin{center}\includegraphics[clip=true,angle=0,width=0.75\textwidth]{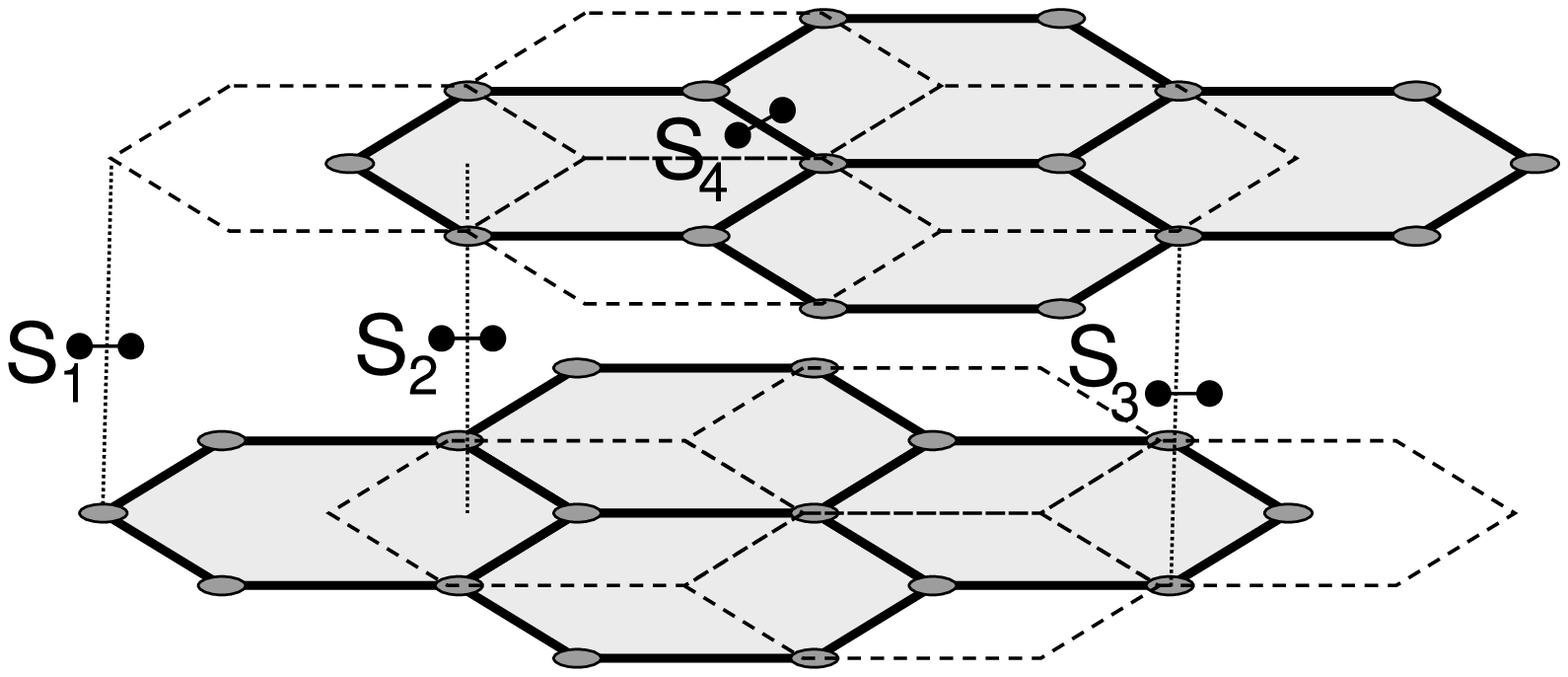}\end{center}
\caption{ $H_2$ molecule at the symmetry sites (..AB.. stacking).
$S_1$:$H_2$ molecule is over the carbon atom.
$S_2$:$H_2$ molecule is inside an empty hexagonal site.
$S_3$:$H_2$ molecule is between two carbon atoms.
$S_4$:$H_2$ molecule is over the bridge.
In all these configurations the $H_2$ molecule axis is held parallel to the
layer.}
\label{fig-1A}
\end{figure}

To investigate hydrogen adsorption in graphite systems, we first performed test calculations on
pure graphite.
In excellent agreement with previous calculations, we found the in-plane C-C bonds to have
a length of 1.42 \AA{}, while the c-axis -- {\it i.e.,} the distance between the graphene
sheets was found to be 3.35 \AA{}, respectively. Next, we tested the energetics of adding a
single $H_2$ molecule to the various symmetry sites. These different sites are shown
in Fig. \ref{fig-1A}, while their energetics is summarized in Table \ref{table1}. 
For the pure graphitic system, the most favored site $S_1$ lies above a carbon atom, midway 
between two planar graphene sheets. Because of the ..ABAB.. stacking of the graphene sheets, other
symmetry sites such as $S_2$, which is located inside an "empty" hexagonal 
site, $S_3$ and $S_4$, are somewhat less favorable, indicating that the hydrogen molecule is unlikely to
insert into these positions.

Now consider hydrogen in Li-doped graphitic systems. Here, we focus on stage one intercalation
with a Li uptake of one Li atom for every six carbon atoms, {\it i.e.}, $LiC_6$, 
since this configuration should lead to maximal hydrogen uptake. 
The presence of Li atoms is known to change the stacking of the graphene sheets from ..ABAB.. to ..AAAA.. 
In addition, Li transfers its valence electron completely to the graphene sheets. Figure 2
represents the structure of $LiC_6$, with the location of the different symmetry sites marked.

\begin{figure}[htbc]
\begin{center}\includegraphics[clip=true,angle=270,width=0.5\textwidth]{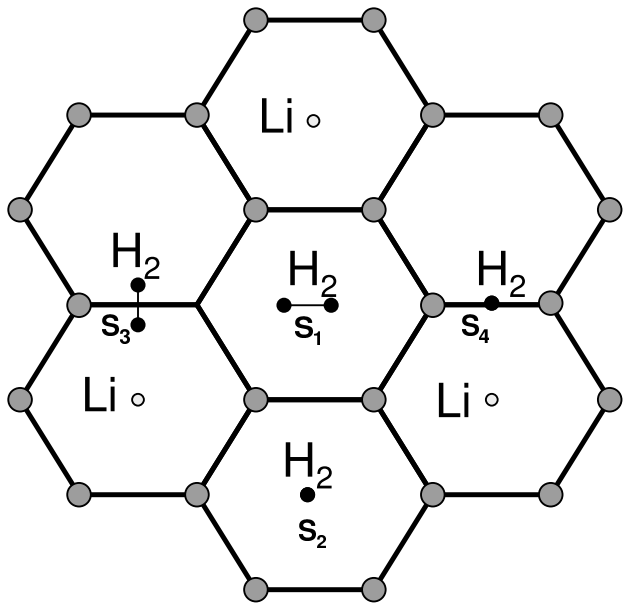}\end{center}
\caption{Energetically favorable configurations for $H_2$ molecule approach to
the graphene layer in Li doped graphite (..AA.. stacking, for view).
$S_1$: $H_2$ molecule, at the center of hexagon, parallel to layer.
$S_2$: $H_2$ molecule, at the center of hexagon, orthogonal to layer.
$S_3$: $H_2$ molecule, over C-C bond, parallel to layer.
$S_4$: $H_2$ molecule, over C-C bond, orthogonal to layer.}
\label{fig-2A}
\end{figure}

Table 1 summarizes the adsorption energies of $H_2$ at these sites. The minimum energy
site in this case is located at $S_1$, which is located inside the "empty" hexagonal
cells, midway between any two graphene layers. All other sites cost at least 0.2
eV more in energy. In all cases, the ideal configuration of the molecular hydrogen 
corresponds to the molecular axis being parallel to the graphene sheet as shown in
Fig. \ref{fig-2A} . As expected, the adsorption is a function of the molecular orientation, with 
other orientations of the hydrogen molecule being less favorable. For instance, for a hydrogen
molecula at $S_2$ whose molecular axis is perpendicular to the graphene sheet is
0.22 eV less favorable. Similarly, an orientation with the molecular axis  around 45$^0$
is less favorable by about 0.5 eV.

\begin{figure}[htbc]
\begin{center}\includegraphics[clip=true,angle=270,width=0.5\textwidth]{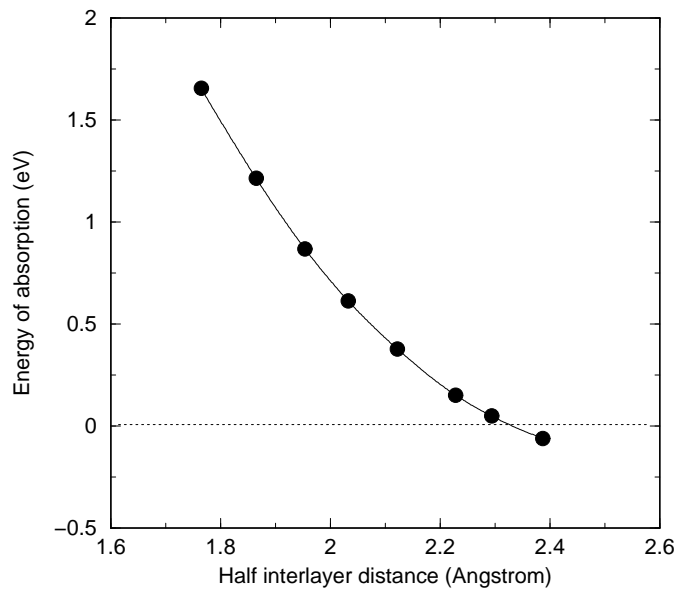}\end{center}
\caption{Energy of absorption for LiC$_6$H$_2$ unit cell (..AA.. stacking).
The curve is not a fit, but simply a guide to the eye. }
\label{fig-3A}
\end{figure}

Having identified the ideal adsorption sites, we turn to the thermodynamics of hydrogen
uptake. At low temperatures, the chemical potential $\mu$ is well approximated by
$\mu =E_{System+H_2}-E_{System}-E_{H_2}$, where quantities like $E_{system+H2}$ represent
the total energy of the system with an adsorbed hydrogen molecule. For a graphite system,
we obtain $\mu = 2.3$ eV, in agreement with previous investigations\cite{Jacobson}. For 
Li-intercalated graphite, we obtain $\mu =0.144$ eV for hydrogen adsorption at the most 
favorable sites. Clearly, since these values are positive, hydrogen uptake is not 
thermodynamically favored for either of the two material systems.
What is the origin of the large change in the chemical potential? It is largely due to the
change in the ABAB to AAAA stacking, and a swelling of the Li-intercalated compound.
For instance for Li-intercalated graphite, the addition of a single hydrogen molecule
per unit cell causes a dramatic swelling of the distance between the graphene
sheets, we note at least a substantial 24 percent change. To provide further evidence of this
effect, we have studied the change in chemical potential as a function of distance between
the graphene sheets. We find that a very large distance of 4.6 \AA{} is
required before $\mu$ becomes nearly zero (Fig. \ref{fig-3A}). Such a large change implies that it very 
unlikely that there is any significant hydrogen uptake in either of these two graphite-based systems.

\smallskip
\noindent
{\bf (c) $H_2$ in doped carbon nanotube ropes}

Having discussed hydrogen adsorption in Li-doped graphite, we turn to adsorption
in Li-doped nanotube systems. Nanotube ropes consist of bundles of nanotubes whose
centers are arranged in a hexagonal pattern. $Li$ ions intercalate into the ropes
primarily in the channels between the nanotubes with a {\it 1:6} ratio. Associated with
the Li intercalation is a slight swelling of carbon nanotube ropes, which changes
the distance between the nanotubes from 3.3 to 3.42 \AA{}. Now, in its {\it local}
structure, carbon nanotubes are very similar to planar graphite sheets, so that in
terms of hydrogen adsorption we can expect very similar results. To test
this we constructed a prototypical nanotube rope out of (5,5) armchair nanotubes.
Inside this rope, we placed 6 Li atoms per unit cell; each unit cell contained 40
carbon atoms. 
Hydrogen molecules were then added to the system as in the case of Li-intercalated 
graphite and the chemical potential changes were computed as already described. 
The results are similar to those obtained for the graphite system: only by greatly increasing
the distance between nanotubes does it become thermodynamically favorable for hydrogen 
molecules to adsorb into the system. Hence, similar to the case of hydrogen adsorption
into the interstitial channels of nanotube ropes, we do not expect significant amounts
of hydrogen molecules to be stored in the corresponding Li-intercalated system.
For instance, in our estimate, we could put at most 2 $H_2$ molecules for a (10,10)
nanotube at the interstices. 

What about hydrogen molecules entering nanotube interiors? Clearly, hydrogen molecules
must be able to diffuse into and out of these interior regions at a reasonable rate
if carbon nanotube systems are to act as a convenient molecular container. To probe
the ability of hydrogen molecules to enter a nanotube interior, we calculated the 
energetics of a diffusion pathway for the molecule to move through the sidewall of a (5,5)
nanotube. While there are diffusion minima on either side of the nanotube wall, a very
large barrier of about 10.72 eV is encountered, so that it is very unlikely that hydrogen
molecules can diffuse into nanotubes through pristine sidewalls. Hence, other pathways
into the nanotubes must be found. A possible option here is provided by the various
topological defects that can be induced in the nanotube walls by such nonequilibrium
means such as ion bombardment \cite{Pajayan}, tensile straining, or by the mechanical 
grinding of nanotubes \cite{Nardelli,Orlikowski}. The spirit here is similar to previous work on 
Li-intercalation into nanotube systems. While initial experimental results showed a 
relatively poor 25$\%$ increase of Li uptake relative to graphite, better results were 
achieved by first ball-milling \cite{Gao} the carbon nanotube samples. This not only dramatically 
reduced the irreversible Li capacity of nanotubes, but also increased the reversible capacity 
to about 2.7 Li atoms per 6 C atoms.

One can expect that the structure of nanotube defects to vary considerably, depending on
how they are induced. However, the relevant feature of all the various defects is that
they consist of n-membered rings in various combinations. Hence, we focused on the
diffusion barriers for hydrogen molecules such rings, as shown in Fig. \ref{fig-5A}. Note that
once formed, we can expect these metastable structures to persist for very long periods of
time because of the high activation energies required for bond rotations\cite{QZhao}. 
Table \ref{table2} summarizes the diffusion barriers for hydrogen molecules to enter nanotube interiors. 

As expected, as the number of sides increases and the "holes" become larger and larger, the diffusion
barriers drop. However, even for 10-sided rings or decagons, a significant barrier of
1.6 eV in encountered. Hence, defects larger than decagons are required for
efficient hydrogen motion in and out of nanotubes interiors.
These results are quite different from the case of Li ions, for which the diffusion
barriers are all about 1-2 eV lower indicating that enhanced Li uptake through defects
is indeed possible\cite{Chris-Vincent}. 

\begin{figure}[htbc]
\begin{center}\includegraphics[clip=true,angle=270,width=0.5\textwidth]{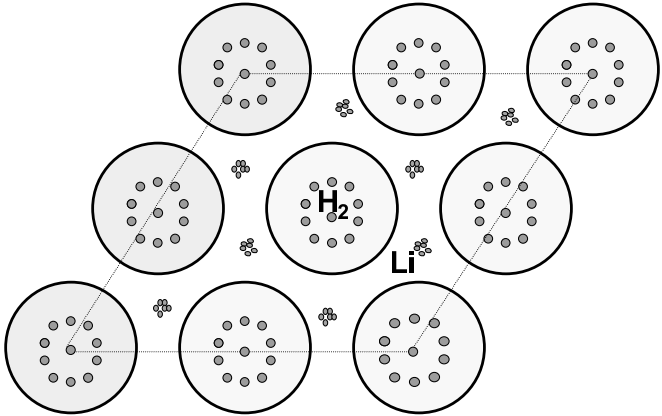}\end{center}
\caption{Top projected view of hydrogen and lithium intercalation in nanotube
ropes. Lithium atoms (not coplanar) are in the interstitial channels while $H_2$ molecules are in the
nanotube interiors.}
\label{fig-4A}
\end{figure}

\begin{figure}[htbc]  
\includegraphics[clip=true,angle=90,width=0.75\textwidth]{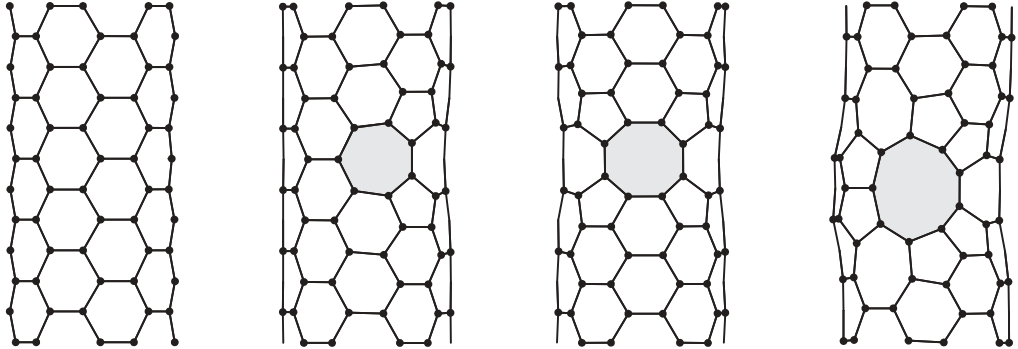}
\caption{ Structure of the topological defects displaying different {\it n-}membered rings. 
The diffusion barriers through these rings and the corresponding formation energies 
are given in Table II.}
\label{fig-5A}
\end{figure}

Finally, the only other way in which hydrogen molecules can enter nanotube interiors is
via their open ends, provided that such tubes can be stabilized. To be useful, such open-ended
tubes would have to be relatively short in order to allow for a reasonable rate of
molecular exchange. To place a hydrogen molecule inside an open nanotube, we find that there is
essentially no diffusion barrier. In fact, there is a small net decrease in the chemical
potential of the system of about 0.16 eV, indicating that a single hydrogen molecule is
readily pulled into the nanotube interior. Similarly, a second hydrogen molecule may
be added to a unit cell consisting of 4 rings of a (5,5) nanotube (40 atoms) at virtually 
no cost in in terms of a chemical potential. However, we find that
adding a third molecule requires a $\mu = 1.09$ eV. Hence, a (5,5) nanotube interior will
only hold 2 hydrogen molecules for every 4 rings giving a relatively small ratio
of 0.76 weight percent. Since larger diameter nanotubes should in principle be able to
accommodate more hydrogen molecules, based on geometry, we performed both LDA and GGA calculations 
for an Li doped (10,10) nanotube ({\it 1:6} ratio) in order to assess the size (diameter) 
dependence of the nanotubes on the hydrogen to carbon ratio. 
We obtained 3.33 wt $\%$ and 2.24 wt $\%$ (the wt $\%$ from the interstitial
sites was included) for hydrogen intake using LDA and GGA respectively. 
In the LDA calculation, the nanotube interior held 16 hydrogen molecules per 80
carbon atoms arranged in two heptagonal layers, whereas for GGA the configuration was 
pentagonal; finally, in the LDA case only, among the 16 molecules, 2 hydrogen molecules were
located along the nanototube axis as shown in Fig. \ref{fig-4A}. While there is a net increase
compared to a (5,5) tube, the hydrogen to carbon ratio is not boosted sufficiently. Thus, even if 
nanotube interiors are somehow made accessible, one can still only expect relatively small uptake
of hydrogen molecules.
  
\bigskip
\centerline
{\bf III. Summary}

We have investigated the uptake of hydrogen molecules in Li-doped graphite and carbon nanotube
systems with {\it ab initio} simulations. For both systems, there is no evidence that
doping with alkali metals should lead to significant increases in the adsorption of molecular
hydrogen\cite{Meregalli,Froudakis}. Molecular hydrogen does not readily
intercalate into pure a graphite, unless the distance between the graphene sheets 
is artifically enlarged. Similar results hold for the Li-doped graphite and carbon nanotube ropes,
although the chemical potential needed to add a hydrogen molecule is significantly reduced.
Our theoretical results are therefore not consistent with the original experiment of 
Chen et al\cite{7}. We speculate that a suitable explanation for this discrepancy has been
recently found by Pinkerton and coworkers\cite{Pinkerton}, who attribute the enhanced hydrogen
uptake in the original experiment to water contamination. In addition,
we have studied the diffusion barriers for hydrogen molecules to enter into nanotube
interiors. Because of the very large diffusion barriers, molecular hydrogen cannot enter through
nanotube sidewalls. Preprocessing nanotube samples by mechanical means 
such as grinding will require the introduction of large-ring topological defects
(greater than decagons) in order to enable access to nanotube interiors.
However, even if nanotubes interiors are accessible, the hydrogen uptake is
still modest.

\smallskip

We thank Vincent Meunier for numerous scientific discussions and the {\it ab initio}
coordinates of the nanotube defects. This work was supported both by NASA and ONR grants. 
We also thank the North Carolina Supercomputing Center (NCSC) for
extensive computer support.

\begin{table}[htbc]
\caption{Energies (eV) for $H_2$ absorption sites, measured with respect to the
value at $S_1$.}
\begin{ruledtabular}
\begin{tabular}{lcccc}
&$S_1$& $S_2$& $S_3$& $S_4$\\
\hline
Carbon  & 0& 0.19& 0.38& 0.76\\
Carbon-Li  & 0& 0.22& 0.46& 0.65\\
\end{tabular}
\end{ruledtabular}
\label{table1}
\end{table}

\begin{table}[htbc]
\caption{Diffusion barriers for hydrogen to move through a given n-membered
ring.}
\begin{ruledtabular}
\begin{tabular}{lccccc}
N-sided rings&Hexagon&Heptagon&Octagon&Enneagon&Decagon\\
\hline
$\Delta E$ (eV)& 10.72 & 7.75 & 5.68 & 2.23 & 1.63 \\
Formation (eV)& 0.0 & 3.5  & 6.2  & 9.5  & -\\ 
\end{tabular}
\end{ruledtabular}
\label{table2}
\end{table}

\end{document}